\documentclass[twocolumn,superscriptaddress,nofootinbib,aps,prl,floatfix,preprintnumbers,amsmath,amssymb,groupedaddress]{revtex4}

\usepackage{dsfont}
\usepackage{epsfig}
\usepackage{slashed}
\usepackage{bbold}
\usepackage{psfrag}
\usepackage{color}
\PassOptionsToPackage{caption=false}{subfig}
\usepackage{subfig}
\usepackage{multirow}
\usepackage{booktabs}
\PassOptionsToPackage{hyphens}{url}
\usepackage{hyperref}
\begin{document}

\def\as{\alpha_S}
\def\t{{\bar t}}
\def\dy{{\Delta y}}
\def\dmody{{\vert\Delta y\vert}}
\def\Mtt{M_{t\bar t}}
\def\PT{P_{T,t\bar t}}
\def\GeV{\, \rm GeV}
\def\AFB{\rm A_{FB}}
\def\AFBincNUM{0.095 \pm 0.007}
\title{Resolving the Tevatron Top Quark Forward-Backward Asymmetry Puzzle:\\ Fully Differential Next-to-Next-to-Leading-Order Calculation}
\author{Michal Czakon}
\affiliation{Institut f\"ur Theoretische Teilchenphysik und Kosmologie,
RWTH Aachen University, D-52056 Aachen, Germany}

\author{Paul Fiedler}
\affiliation{Institut f\"ur Theoretische Teilchenphysik und Kosmologie,
RWTH Aachen University, D-52056 Aachen, Germany}

\author{Alexander Mitov}
\affiliation{Cavendish Laboratory, University of Cambridge, Cambridge CB3 0HE, UK}

\preprint{Cavendish-HEP-14/10, TTK-14-32}

\begin{abstract}
We determine the dominant missing Standard Model (SM) contribution to the top quark pair forward-backward asymmetry at the Tevatron. Contrary to past expectations, we find a large, around $27\%$, shift relative to the well-known value of the inclusive asymmetry in next--to--leading order (NLO) QCD. Combining all known Standard Model corrections, we find that $\AFB^{\rm SM} = \AFBincNUM$. This value is in agreement with the latest D\O\ measurement [V. M. Abazov {\it et al.} (D0 Collaboration), Phys. Rev. D 90, 072011 (2014)] $\AFB^{\rm D\O} = 0.106\pm 0.03$ and about $1.5\sigma$ below that of CDF [T. Aaltonen {\it et al.} (CDF Collaboration), Phys. Rev. D 87, 092002 (2013)] $\AFB^{\rm CDF} =0.164\pm 0.047$. Our result is derived from a fully differential calculation of the next--to--next--to leading order (NNLO) QCD corrections to inclusive top pair production at hadron colliders and includes -- without any approximation -- all partonic channels contributing to this process. This is the first complete fully differential calculation in NNLO QCD of a two--to--two scattering process with all coloured partons. 
\end{abstract}
\maketitle

\section{Introduction}

At the Tevatron $p\bar p$ collider top quarks are produced predominantly in the hemisphere defined by the direction of the proton beam \cite{Kuhn:1998jr,Kuhn:1998kw}. Such a production rate difference is often referred to as Forward-Backward Asymmetry ($\AFB$). The Tevatron collider is uniquely positioned for the measurement of this asymmetry since $\AFB$ is not present at $pp$ colliders, e.g.\ the LHC (although a related, albeit strongly diluted asymmetry can be measured at the LHC; see for example \cite{AguilarSaavedra:2012rx} for more details).

This unique Tevatron capability, coupled with the persistent discrepancy \cite{Aaltonen:2011kc} between the measured and predicted $\AFB$, have turned this observable into one of the most influential measurements performed at the Tevatron. Indeed, the $\AFB$-related publications by the CDF \cite{Aaltonen:2008hc,Aaltonen:2011kc,Aaltonen:2012it,CDF:2013gna,Aaltonen:2013vaf,Aaltonen:2014eva} and D\O\ \cite{Abazov:2007ab,Abazov:2011rq,Abazov:2012oxa,Abazov:2013wxa,Abazov:2014oea,Abazov:2014cca} collaborations have initiated major research activity both in explaining the discrepancy with beyond the Standard Model (BSM) physics (see e.g. Refs.~\cite{Kamenik:2011wt,Berger:2011ua}) and in estimating $\AFB$ within the Standard Model \cite{Kuhn:1998jr,Kuhn:1998kw,Antunano:2007da,Almeida:2008ug,Kidonakis:2011zn,Ahrens:2011uf,Hollik:2011ps,Kuhn:2011ri,Manohar:2012rs,Campbell:2012uf,Brodsky:2012ik,Skands:2012mm,Bernreuther:2012sx} (see Ref.~\cite{Aguilar-Saavedra:2014kpa} for an in-depth review).

The effort to reconcile this discrepancy within the SM has so far been hampered because of the lack of a convincing estimate of the missing SM corrections. In this work we calculate the dominant missing correction and provide a realistic uncertainty estimate for $\AFB$ in the SM. Our conclusion is that the SM prediction is under good theoretical control and agrees very well with the latest measurement -- both inclusive and differential -- from the D\O\ \cite{Abazov:2014cca} collaboration. For inclusive $\AFB$, we find reasonable agreement with the latest measurement from the CDF collaboration \cite{Aaltonen:2012it}.

\section{$\boldsymbol{\AFB}$: brief history and current status}\label{sec:AFB-status}

The focus of this work is $\AFB$ for stable top quarks. For lepton-level $\AFB$, we refer the reader to Refs.~\cite{Falkowski:2012cu,Campbell:2012uf,Bernreuther:2012sx, Abazov:2012oxa,Aaltonen:2013vaf,Abazov:2013wxa,Abazov:2014oea,Aaltonen:2014eva}.

A non-vanishing $\AFB$ is predicted at next-to-leading order (NLO) in QCD. It was originally evaluated by K\"uhn and Rodrigo \cite{Kuhn:1998jr,Kuhn:1998kw} long before the first measurements became available.
The early measurements of $\AFB$ showed \cite{Aaltonen:2011kc} a very large discrepancy with respect to the SM prediction, especially at large $t\t$ invariant mass $\Mtt>450~\GeV$. Subsequent refinements of the measurements established \cite{Aaltonen:2012it} a less-pronounced $\AFB$ at large $\Mtt$, which was still $2\sigma$ to $3\sigma$ above the SM prediction. Earlier this year, the D\O\ collaboration published \cite{Abazov:2014cca} an $\AFB$ measurement at full data set, which turned out to be significantly lower than that of CDF \cite{Aaltonen:2012it} and thus much closer to the SM predictions. 

The significance of the discrepancy between measurement and the SM theory prediction for $\AFB$  has always critically hinged on the size of missing higher-order corrections. Here, we recall the calculation of the NLO QCD corrections \cite{Dittmaier:2007wz} to $\AFB$ in the related process $t\t j$, where a nearly $-100\%$ correction was found. Such a very large correction, if it were to also appear in $t\t$, would have had the potential of removing the discrepancy. Still, a careful analysis performed by Melnikov and Schulze \cite{Melnikov:2010iu} suggests that $\AFB$ in $t\t$ is unlikely to receive very large corrections in the next order in QCD (i.e. in NNLO QCD) and is ``{\it most likely stable against yet higher order corrections}''. Our calculation of the NNLO QCD correction to $\AFB$ is in line with their findings. (We equate ``large'' with ``important, but not spoiling perturbative convergence'', while ``very large'' might imply spoiling of perturbative convergence).

In a series of papers \cite{Hollik:2011ps,Kuhn:2011ri,Bernreuther:2012sx} it was found that, unexpectedly, electroweak (EW) corrections to $\AFB$ are quite large. For example, for inclusive $\AFB$, they are around $25\%$ of the NLO QCD term. Contributions from Sudakov EW corrections have also been computed \cite{Manohar:2012rs}.

So far, the only source of information about higher-order QCD corrections to $\AFB$ has been soft-gluon resummation. It was first applied at next-to-leading logarithmic accuracy (NLL) in Ref.~\cite{Almeida:2008ug} and later extended to NNLL in Ref.~\cite{Kidonakis:2011zn,Ahrens:2011uf}. Further understanding of the nature of such soft emissions came in the context of parton showers and from probing them down to a single gluon emission \cite{Skands:2012mm}. From Refs.~\cite{Almeida:2008ug,Ahrens:2011uf,Skands:2012mm} one concludes that, beyond NLO QCD, soft-gluon emission generates negligible corrections to inclusive $\AFB$. The natural interpretation of this result, especially when augmented with the conclusions of Ref.~\cite{Melnikov:2010iu}, was that the missing NNLO QCD contributions to $\AFB$ in $t\t$ may be small and may not significantly affect the SM $\AFB$ prediction. Contrary to the above expectations we find that the NNLO QCD corrections are large and originate mostly from emissions that are not controlled by soft-gluon resummation. 

An alternative approach to computing $\AFB$, based on the Principle of Maximum Conformality (PMC) \cite{Brodsky:2012rj} scale setting, was used in Ref.~\cite{Brodsky:2012ik}. The authors derive a value for $\AFB$, which is significantly higher than the usual NLO QCD correction, in agreement with the CDF measurement. While the related Brodsky-Lepage-Mackenzie (BLM) \cite{Brodsky:1982gc} scale setting procedure is known \cite{Melnikov:2005bx} to work well even beyond fully inclusive observables, its applicability in top production at hadron colliders is not as established. For example, the NNLO results \cite{Czakon:2013goa,Czakon:2012pz,Czakon:2012zr,Baernreuther:2012ws} for the terms quadratic in the number of massless quarks ($N_F$) in the total $t\t$ cross-section differ from those predicted within the BLM approach. (In particular, the term $\propto N_F^2$ in $q\bar q\to t\t +X$ is known analytically \cite{Baernreuther:2012ws}. The difference with respect to the BLM prediction is $\propto \pi^2\sigma_{\rm Born}\,,$ and can be thought of as due to an analytical continuation to space-like kinematics).

Finally, we recall the impact on $\AFB$ from asymmetries in the subtracted $t\t$ backgrounds \cite{Hagiwara:2012td}, as well as the possibility \cite{final-state-inter,Abazov:2014cca} that final state $t\bar t$--spectator interactions could contribute to $\AFB$. The latter problem has been addressed in Ref.~\cite{Mitov:2012gt}, where it was shown that such interactions are strongly suppressed for single-inclusive top (or $\t$) observables but need not be for double-inclusive observables (like the ones we study in this paper) in the presence of strong jet vetoes. (The agreement between single- and double-inclusive measurements of $\AFB$ \cite{Aaltonen:2011kc} might be an indication that such a mechanism for generating $\AFB$ in inclusive $t\bar t$ production may not be playing a significant role. Improved modelling of the so-called gap fraction \cite{ATLAS:2012al} may help in clarifying this issue).

\section{Results}

Following \cite{Aaltonen:2012it}, the differential asymmetry is defined as
\begin{equation}
{\AFB} = {\sigma^{+}_{\rm bin} - \sigma^{-}_{\rm bin} \over \sigma^{+}_{\rm bin} + \sigma^{-}_{\rm bin} }
~~~,~~~
\sigma^\pm_{\mathrm{bin}} = \int \theta(\pm \Delta y) \theta_{\mathrm{bin}} \mathrm{d}\sigma\, ,
\label{eq:AFB-diff}
\end{equation}
with the rapidity difference $\dy \equiv y_t - y_\t$. The binning function $\theta_{\mathrm{bin}}$ restricts the kinematics of  the $t\bar t$ pair to the corresponding bins in figs.~\ref{fig:Y},\ref{fig:MTT},\ref{fig:Ptt}. Setting $\theta_{\mathrm{bin}}=1$ in eq.~(\ref{eq:AFB-diff}) yields the inclusive asymmetry $\AFB$.

The fully differential cross-section $\mathrm{d}\sigma$ appearing in eq.~(\ref{eq:AFB-diff}) for the process $p\bar p \to  t\t+X$ is computed through NNLO in the strong coupling $\as$. We use the top pole mass $m_t=173.3\GeV$, the MSTW2008 pdf set \cite{Martin:2009iq} and kinematics-independent scales with central value $\mu_R=\mu_F=m_t$. The theoretical uncertainty is estimated with restricted scale variation $\mu_R\neq\mu_F\in (m_t/2,2m_t)$ \cite{Cacciari:2008zb} which was validated with the NNLO $t\t$ cross-section \cite{Czakon:2013goa,Czakon:2012pz,Czakon:2012zr,Baernreuther:2012ws}. The pdf uncertainty is small and is not included. 

The differential cross-section $\mathrm{d}\sigma$ is computed following the setup of Refs.~\cite{Czakon:2013goa,Czakon:2012pz,Czakon:2012zr,Baernreuther:2012ws}: the two-loop virtual corrections are evaluated as in Refs.~\cite{Czakon:2008zk,Baernreuther:2013caa}, utilising the analytical form for the poles \cite{Ferroglia:2009ii}. The one-loop squared amplitude has been calculated previously \cite{Anastasiou:2008vd} and confirmed by us. The real-virtual (RV) corrections are derived by integrating the one-loop amplitude with a counter-term that regulates all its singular limits \cite{Bern:1999ry}. The finite part of the one-loop amplitude is computed with a code used in the calculation of $pp \to t\t j$ at NLO \cite{Dittmaier:2007wz}. The double real corrections (RR) are computed as in Refs.~\cite{Czakon:2010td,Czakon:2011ve}. 

Our calculation includes {\it all} partonic reactions that contribute to inclusive $t\t$ production in pure QCD without making any approximations. We have checked that our calculation reproduces $\sigma_{\rm tot}$ from Refs.~\cite{Czakon:2013goa,Czakon:2012pz,Czakon:2012zr,Baernreuther:2012ws} for each value of $\mu_R,\mu_F$ with a precision better than one permil. We also observe the cancellation of infrared singularities in each bin. At NLO our calculation agrees with the MCFM Monte Carlo generator \cite{Campbell:2012uf,Nason:1987xz}. The predicted NNLO $\PT$ dependence of $\AFB$ for non-vanishing transverse momentum, $\PT\geq 10\GeV$ (see fig.~\ref{fig:Ptt}), is consistent with results for the NLO QCD corrections to $pp \to t\t j$ from Refs.~ \cite{Melnikov:2010iu,Melnikov:2011qx,Hoeche:2013mua} and agrees perfectly with an independent evaluation using \textsc{Helac-Nlo} \cite{Bevilacqua:2011xh}.
  
In this work we use two definitions for $\AFB$ that are formally equivalent through NNLO and allow for EW corrections
\begin{eqnarray}
{\AFB} &\equiv& { N_{\mathrm{EW}} + \as^3N_3 + \as^4N_4 + {\cal O}(\as^5) \over \as^2D_2+\as^3D_3+\as^4D_4+{\cal O}(\as^5)} \label{eq:AFB-no-expand}\\
&=& \as{N_3\over D_2} + {N_{\mathrm{EW}}\over \as^2D_2}  \label{eq:AFB-expand}\\ \nonumber
&& + \as^2\left({N_4\over D_2}-{N_3D_3\over D_2^2} \right) - {N_{\mathrm{EW}}D_3\over \as D_2^2} +  {\cal O}(\as^3) \, .
\end{eqnarray}
[The term $N_{\mathrm{EW}}$ contains some terms that involve powers of $\as$. We ignore this $\as$-dependence in the power counting in eq.~(\ref{eq:AFB-expand}).] The first definition, eq.~(\ref{eq:AFB-no-expand}), uses exact results in both numerator and denominator of eq.~(\ref{eq:AFB-diff}), while the second, eq.~(\ref{eq:AFB-expand}), is the expansion of the ratio eq.~(\ref{eq:AFB-no-expand}) in powers of $\as$. (Such an expansion is not, strictly speaking, fully consistent since the $\as$ expansion is performed after convolution with pdf's. Nevertheless, following the existing literature, we consider it as an indication of the sensitivity of $\AFB$ to missing higher order terms.)

\begin{figure}[t]
\centering
\hspace{0mm} 
\includegraphics[width=0.49\textwidth]{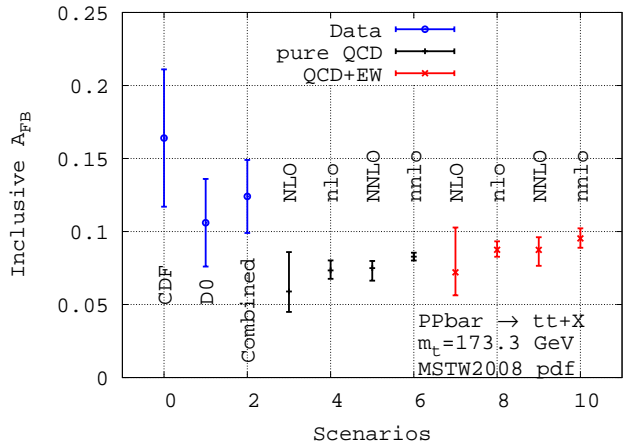} 
\caption{The inclusive asymmetry in pure QCD (black) and QCD+EW\cite{Bernreuther:2012sx} (red). Capital letters (NLO, NNLO) correspond to the unexpanded definition (\ref{eq:AFB-no-expand}), while small letters (nlo, nnlo) to the definition (\ref{eq:AFB-expand}).
The CDF/D\O\ (naive) average is from Ref.~\cite{Aguilar-Saavedra:2014kpa}. Error bands are from scale variation only. Our final prediction corresponds to scenario 10.}
\label{fig:A-inclusive}
\end{figure}

\begin{figure}[t]
\centering
\hspace{0mm} 
\includegraphics[width=0.49\textwidth]{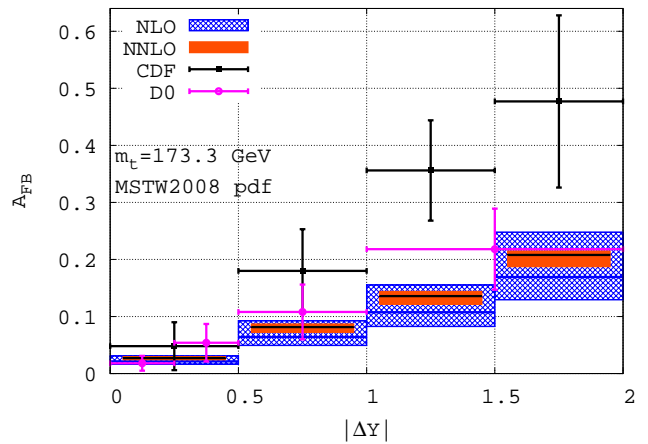} 
\caption{The $\dmody$ differential asymmetry in pure QCD at NLO (blue) and NNLO (orange) versus CDF \cite{Aaltonen:2012it} and D\O\ \cite{Abazov:2014cca,D0:public} data. Error bands are from scale variation only. For improved readability some bins are plotted slightly narrower. The highest bin contains overflow events.}
\label{fig:Y}
\end{figure}

\begin{figure}[h]
\centering
\hspace{0mm} 
\includegraphics[width=0.49\textwidth]{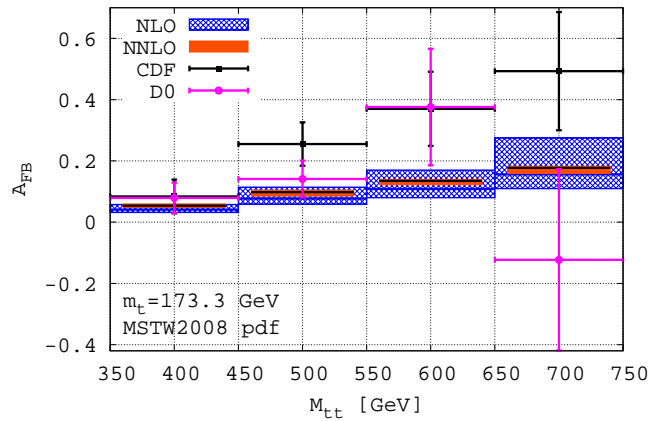} 
\caption{As in fig.~\ref{fig:Y} but for the $\Mtt$ differential asymmetry. The highest bin contains overflow events and the lowest bin includes all events down to the production threshold $2m_t$.}
\label{fig:MTT}
\end{figure}

\begin{figure}[h]
\centering
\hspace{0mm} 
\includegraphics[width=0.49\textwidth]{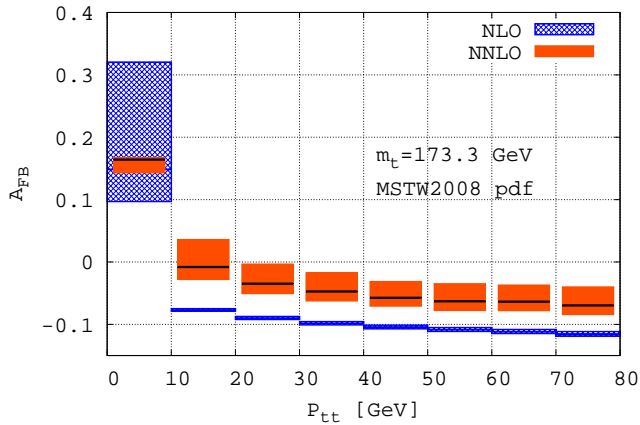} 
\caption{As in fig.~\ref{fig:Y} but for the $\PT$ differential asymmetry.}
\label{fig:Ptt}
\end{figure}

In the present letter, we present differential asymmetries with the unexpanded definition (\ref{eq:AFB-no-expand}) and without EW corrections (see figs.~\ref{fig:Y},\ref{fig:MTT},\ref{fig:Ptt}). The inclusive asymmetry, see fig.~\ref{fig:A-inclusive}, is computed with both definitions (\ref{eq:AFB-no-expand}) and (\ref{eq:AFB-expand}) including EW corrections. (EW corrections to $D_i$ are neglected since EW effects to the total cross-section are very small ${\cal O}(1\%)$, see Refs.~\cite{Beenakker:1993yr,Moretti:2006nf,Bernreuther:2006vg,Hollik:2007sw,Kuhn:2013zoa}.) The numerator factor $N_{\mathrm{EW}}$ is taken from Table 2 in Ref.~\cite{Bernreuther:2012sx}. (We have checked that the different pdf and $m_t$ used in Ref.~\cite{Bernreuther:2012sx} have negligible impact on the QCD numerator $N_3$ and so we expect the same to hold for $N_{\mathrm{EW}}$.) Only for the inclusive asymmetry we determine the scale variation by keeping $\mu_R=\mu_F$ (since the scale dependence of $N_{\mathrm{EW}}$ is published \cite{Bernreuther:2012sx} only for $\mu_R=\mu_F$). (We have checked that for the pure QCD corrections to the total asymmetry the difference with respect to scale uncertainty derived with $\mu_R \neq \mu_F$ variation is negligible.) We also note that the scale variation of $\AFB$ is derived from the consistent scale variation of the ratio, i.e. both numerator and denominator in eqs.~(\ref{eq:AFB-no-expand}) and (\ref{eq:AFB-expand}) are computed for each scale value.


\section{Discussion and conclusions}

In fig.~\ref{fig:A-inclusive} we observe that the central values of the expanded (\ref{eq:AFB-expand}) and unexpanded (\ref{eq:AFB-no-expand}) definitions of inclusive $\AFB$ differ significantly at NLO but less so at NNLO. While the unexpanded definition (\ref{eq:AFB-no-expand}) closely resembles the experimental setup, the consistency of the two definitions {\it within uncertainties} renders the question about the more appropriate choice largely irrelevant.
We also note the small scale error for the expanded $\AFB$ definition (\ref{eq:AFB-expand}) in pure QCD at both NLO and NNLO, which appears too small to be realistic. The inclusion of EW corrections, however, breaks this pattern and brings the scale dependence in line with the unexpanded definition eq.~(\ref{eq:AFB-no-expand}). 
Therefore, following the previous literature, we choose as our final prediction $\AFB^{\rm SM}=\AFBincNUM$ (scenario 10 in fig.~\ref{fig:A-inclusive}) which is derived with the expanded definition (\ref{eq:AFB-expand}) and includes EW \cite{Bernreuther:2012sx} corrections.

The inclusion of higher order QCD corrections reduces the scale uncertainty of the differential asymmetry. The only exception is the $\PT$ dependent asymmetry whose scale behaviour at NLO QCD is atypical.

\begin{table}[h]
\begin{center}
\begin{tabular}{|c|c|c|c|c|}
\hline
& {\rm Factorization}  & {\rm RR}  & {\rm RV} & {\rm VV} \\
\hline
$({\rm princ.~contr.})/(\as^4N_4$) & $-0.47$ & $5.34$ & $-3.90$ & $0.03$ \\
\hline
\end{tabular}
\caption{\small  Principal contributions to the numerator $N_4$.}
\label{tab:contributions}
\end{center}
\end{table}
The relative contributions of the principal NNLO corrections to the inclusive numerator in eq.~(\ref{eq:AFB-no-expand}) are given in table \ref{tab:contributions}. (Note that this separation is not unambiguous, just as at NLO.) Clearly, the inclusive asymmetry at NNLO is driven by a strong cancellation between RR and RV contributions. The contribution from collinear factorisation is sizeable while the pure virtual (VV) correction is quite small. We have also checked that the numerator $\as^4N_4$ almost exclusively originates in the $q\bar q$ partonic channel. (The contribution due to collinear factorisation is not included in this comparison.) Where present, the contribution to $\as^4N_4$ due to the $qg$ reaction is two orders of magnitude smaller than $q\bar q$. The remaining $qq'$-type partonic reactions are another two orders of magnitude smaller. This pattern is in line with the contributions of these partonic reactions to the total cross-section \cite{Czakon:2013goa,Czakon:2012pz,Czakon:2012zr,Baernreuther:2012ws}.

\begin{table}[h]
\begin{center}
\begin{tabular}{|c|c|c|c|}
\hline
  & NLO  & NNLO & NLO+NNLL \\
\hline
$\as^3N_3 + \as^4N_4$ [pb] & $0.394^{+0.211}_{-0.127}$ & $0.525^{+0.055}_{-0.085}$ &  $0.448^{+0.080}_{-0.071}$  \\
\hline
$\as^4N_4$ [pb] & -- & $0.148$ &  --  \\
\hline
$\AFB$[\%] (eq.~(\ref{eq:AFB-expand})) & $7.34^{+0.68}_{-0.58}$ & $8.28^{+0.27}_{-0.26}$ &  $7.24^{+1.04}_{-0.67}$ \\
\hline
$\AFB$[\%] (eq.~(\ref{eq:AFB-no-expand})) & $5.89^{+2.70}_{-1.40}$ & $7.49^{+0.49}_{-0.86}$ &  -- \\
\hline
\end{tabular}
\caption{\small Comparison of the numerator in eq.~(\ref{eq:AFB-no-expand}) and the inclusive asymmetry $\AFB$ computed in pure QCD at NLO (with NLO pdf set), NNLO and NLO+NNLL \cite{Ahrens:2011uf}. Only errors from $\mu_F=\mu_R$ scale variation are shown.}
\label{tab:results}
\end{center}
\end{table}

In contrast to the negligible approximate NNLO QCD correction to $\AFB$ implied by soft-gluon resummation \cite{Almeida:2008ug,Ahrens:2011uf}, we find that the exact NNLO QCD correction to the inclusive $\AFB$ is, in fact, large. (We note that the prediction of Ref.~\cite{Kidonakis:2011zn} differs from the one of \cite{Almeida:2008ug,Ahrens:2011uf}, presumably due to different subleading terms.) Specifically, in table~\ref{tab:results} we compare the exact results for $\AFB$ and its numerator [defined as the QCD part of the numerator in eq~(\ref{eq:AFB-no-expand})] through NNLO in QCD, with the NLO+NNLL predictions of Ref.~ \cite{Ahrens:2011uf}. (The settings in both papers are the same, except for a small difference of $0.2\GeV$ in the value of $m_t$ which we neglect.) The ratio $\AFB^{\rm (NNLO)}/\AFB^{\rm (NLO)}$ is $1.27$ ($1.13$) for $\AFB$ defined through eq.~(\ref{eq:AFB-no-expand}) (eq.~(\ref{eq:AFB-expand})). The corresponding ratio for the numerator of the asymmetry is $1.33$, which is even larger than that for $\AFB$. Clearly the corrections to both quantities are significantly different from those of approximate NNLO, which yield $0.99$ for the $\AFB$ and $1.13$ for the numerator ratio. [We refrain from directly comparing differential asymmetries because in this work we define them through eq.~(\ref{eq:AFB-no-expand}) while the ones in Ref.~\cite{Ahrens:2011uf} are defined through eq.~(\ref{eq:AFB-expand}).]

The large difference between $\AFB$ predicted in exact and approximate NNLO can be understood from its $\PT$ dependence. We recall that soft gluon resummation applies to kinematical configurations that resemble those at the Born level; i.e., it should mainly contribute to the small $\PT$ bins. 
As fig.~\ref{fig:Ptt} suggests, harder radiation generates a significant portion of the NNLO corrections. Studying the cumulative differential asymmetry $\AFB$$(\PT\leq\PT^{\it cut})$ and the corresponding cumulative numerator we observe that in the first bin $\PT^{\it cut}\leq 10\GeV$ (where soft gluon resummation should be most relevant) the NLO and NNLO numerators are practically equal, i.e. the $10\%$ shift from NLO to NNLO in the first bin in fig.~\ref{fig:Ptt} is exclusively due to the difference between NLO and NNLO denominators. With the inclusion of the next bins, however, the NLO and NNLO cumulative numerators start to differ quite rapidly. Indeed, about $50\%$ of their difference is generated by the addition of the second bin $\PT^{\it cut}= 20\GeV$.

Analysing the $\PT$ dependence of $\AFB$, the CDF collaboration \cite{Aaltonen:2012it} noted that the discrepancy between data and NLO QCD appears to be independent of $\PT$. It is easy to see from fig.~\ref{fig:Ptt} that the difference between NNLO and NLO corrections to the $\PT$ asymmetry for $\PT\geq 10 \GeV$ follows precisely this pattern and is, furthermore, consistent with the analysis of Ref.~\cite{Gripaios:2013rda}.

The pdf uncertainty is generally small and has not been included in our results. For its estimation, we have first computed $\AFB$ in NLO QCD with a NNLO pdf set (at 68\% CL) and then rescaled it with the appropriate $K$-factor based on central scale values. In inclusive quantities such as the inclusive $\AFB$ and the numerator in eq.~(\ref{eq:AFB-no-expand}), the pdf uncertainty is smaller than the scale uncertainty by a factor of 3 or more. Similarly, the pdf error in the differential asymmetry is typically much smaller than the one from scale variation, although in some bins it can be as large as half the scale error. Therefore, for most $\AFB$-related applications we can envisage, one can safely neglect pdf errors. However, if a precise error estimate is essential, the pdf errors might need to be revisited.

The Monte Carlo (MC) integration error in all our results is insignificant. Specifically, its relative contribution to the inclusive asymmetry and cross-section is at the permil and sub-permil levels, respectively.
The relative MC error in the differential asymmetry is typically below $1\%$ in each bin, with the exception of the largest $\Mtt$ bin and the $60\GeV \leq$$\PT\leq 70\GeV$ bin where it is about $1.5\%$ (for central scales).
 
Finally, we would like to emphasise the connection between the top quark $\AFB$ and the perturbatively generated strange asymmetry of the proton \cite{Catani:2004nc}. For example, the asymmetry--generating diagrams are the same in both cases (compare fig.~1 from Ref.~\cite{Catani:2004nc} with fig.~3a of Ref.~\cite{Kuhn:1998kw}) up to crossing legs from the initial to the final state and setting $m_t$ to zero. In fact, {\it in the absence of other predictions}, one might speculate that our results indicate that the currently unknown four-loop corrections to the space-like splitting functions may bring non-negligible corrections to the perturbatively generated $s,c,b,t$ asymmetries of the proton.

{\bf\it Summary.} We compute the largest missing SM correction to top quark $\AFB$ originating in NNLO QCD. Our calculation includes all contributing partonic channels exactly, which makes it the first-ever complete NNLO fully differential calculation in a process with four coloured partons.  In contrast to previous approximations we observe a significant NNLO correction to $\AFB$ which brings the SM prediction for the inclusive asymmetry in agreement with the measurement of the D\O\ collaboration and about $1.5\sigma$ below the value measured by the CDF collaboration. The predicted differential asymmetry, even without EW corrections, is in agreement with the corresponding D\O\ measurements.

\acknowledgments\vskip5mm
We thank Dante Amidei, Tom Ferbel, Amnon Harel, Regina Demina and Jon Wilson for clarifications about the experimental results. We also thank Stefan Dittmaier for kindly providing us with his code for the evaluation of the one-loop virtual corrections, Markus Schulze for a comparison with the results of Ref.~\cite{Melnikov:2010iu,Melnikov:2011qx}, Jan Winter for a comparison with Ref.~\cite{Hoeche:2013mua}, and Manfred Kraus for a comparison with \textsc{Helac-Nlo}. The work of M.~C. and P.~F. was supported by the German Research Foundation (DFG) via the Sonderforschungsbereich/Transregio SFB/TR-9 ``Computational Particle Physics", and the Heisenberg programme. The work of A.~M. is supported by the UK Science and Technology Facilities Council [grants ST/L002760/1 and ST/K004883/1].

\end{document}